# Computation of the off-axis effective area of the New Hard X-ray Mission modules by means of an analytical approach

D. Spiga and V. Cotroneo

INAF/Osservatorio Astronomico di Brera, Via E. Bianchi 46, 23807 Merate, Italy

**ABSTRACT**

One of the most important parameters determining the sensitivity of X-ray telescopes is their effective area as a function of the X-ray energy. The computation of the effective area of a Wolter-I mirror, with either a single layer or multilayer coating, is a very simple task for a source on-axis at astronomical distance. Indeed, when the source moves off-axis the calculation is more complicated, in particular for new hard X-ray imaging telescopes (NuSTAR, ASTRO-H, NHXM, IXO) beyond 10 keV, that will make use of multilayer coatings to extend the reflectivity band in grazing incidence. Unlike traditional single-layer coatings (in Ir or Au), graded multilayer coatings exhibit an oscillating reflectivity as a function of the incidence angle, which makes the effective area not immediately predictable for a source placed off-axis within the field of view. For this reason, the computation of the off-axis effective area has been so far demanded to ray-tracing codes, able to sample the incidence of photons onto the mirror assembly. Even if this approach should not be disdained, it would be interesting to approach the same problem from an analytical viewpoint. This would speed up and simplify the computation of the effective area as a function of the off-axis angle, a considerable advantage especially whenever the mirror parameters are still to be optimized. In this work we present the application of a novel, analytical formalism to the computation of the off-axis effective area and the grasp of the NHXM optical modules, requiring only the standard routines for the multilayer reflectivity computation.

## 1. INTRODUCTION

The launch of a number of X-ray telescopes[1],[2],[3],[4] is foreseen in the next years, most of them carrying optical modules able to effectively focus X-rays beyond the present limit of 10 keV. This will be made possible not only by the increase of the size of the optics, but also by the adoption of shallow incidence angles and multilayer coatings to extend the reflected bandwidth to hard X-rays. The sensitivity of these focusing X-ray telescopes will be determined, among other things, by the effective area and the concentration capabilities of their hard X-ray optical modules. For this reason, the on-axis optical module effective area as a function of the X-ray energy is considered the basic requirement to a mirror module design. Nevertheless, also the effective area off-axis is an important parameter, because it chiefly determines the *Field of View* (FOV) of the telescope: in fact, it is well known that the effective area in general decreases as an astronomical source moves off-axis, due to both geometrical vignetting effects and to the variation of the incidence angles throughout the mirrors surface, which in turn causes a variation of the mirror reflectivity. The angular diameter at which the effective area – at a given X-ray energy - is halved with respect to the corresponding value on-axis is conventionally defined to be the optics FOV at that energy. Another example of the importance of this kind of computation is represented by the effective area variations that may arise when the optical axis randomly oscillates with respect to telescope axis[5], which was a major problem for the SIMBOL-X telescope,[5] which optics and focal plane were expected to fly onboard two separate spacecrafts. Although to a lesser extent, this problem is still to be accurately studied and considered for X-ray telescopes with long focal lengths, because the optics and the detectors are expected to be connected by an extendible truss (like NHXM and IXO), which is not perfectly rigid. The assessment of the effective area variation is then very important to conceive an optical design that is less affected by this drawback.

While in general the computation of the on-axis effective area of Wolter-I mirror for an astronomical source does not pose problems, the same computation off-axis is in general a more complicated task, especially in the case of multilayer-coated mirrors, which exhibit an oscillating reflectivity as a function of the reflection angle and the X-ray wavelength, $\lambda$. This is the reason why this problem has been so far faced along with ray-tracing methods, aimed at reconstructing each X-ray path throughout the mirror shells assembly, and more specifically at sampling the incidence angles of the X-rays reflected on the reflective surfaces. Such routines are recognized as very effective and accurate, but they are often of complex implementation and require a considerable computation time. Even if this is a viable approach, it is also possible to solve that problem from an analytical viewpoint, i.e., along with analytical formulae[7] to express the effective area as a function of $\lambda$. This method is able to return the effective area off-axis for a Wolter-I mirror shell in double cone approximation with *any* reflective coating, including multilayers. Therefore, it represents a way to perform the off-axis area computation more effectively and quickly, and consequently to speed up the optical design process, which may entail a number of effective area evaluations.

In this paper we show some applications of this analytical approach to the NHXM optical modules. In Sect. 2 we review the formulae we use throughout this paper, adding some new results, especially regarding the mirror grasp. In Sect. 3 we show some applications to the case of NHXM. The results are summarized in Sect. 4.

## 2. OFF-AXIS EFFECTIVE AREA AND GRASP OF WOLTER-I MIRRORS

Even though the longitudinal curvature of the parabolic and hyperbolic segments in a Wolter-I mirror is essential to focus X-rays to a single point, for the computation of the on- and off-axis effective area the double cone approximation can be suited to a good approximation, especially for mirrors with a high f-number, *f#*. This allows simplifying the geometry to be used in computing, e.g., the incidence angles for the two reflections of X-rays. We define, as in Fig. 1, $R_0$ to be the radius of the reflecting surface of the shell at the intersection plane, $R_M$ the maximum radius - at the primary segment end, $R_m$ the minimum radius - at the secondary segment end, $f$ the focal length for a source at infinity, $\alpha_0$ the primary segment slope at the intersection plane, fulfilling the condition $R_0 = f \tan(4\alpha_0)$. $L_1$ and $L_2$ are the segment lengths along the optical axis, in general different from each other. Finally, let $\delta = R_0/D$ be the half-divergence of the source at a distance $D$, and $\theta$ to be its off-axis angle. Clearly, $\delta = 0$ for an astronomical source.

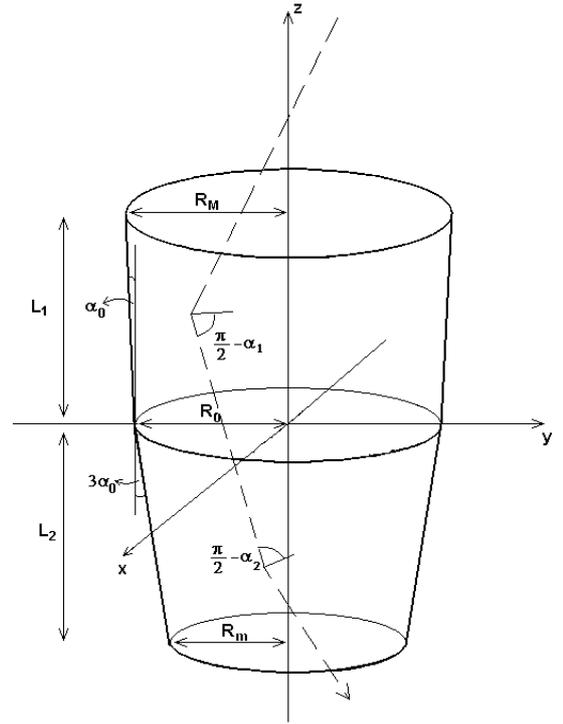

Fig. 1: scheme of a Wolter-I mirror shell. The incidence angles on the primary and secondary segments are shown.

The results of the analytical formalism can be summarized in the following points. For a detailed derivation of the results and the comparison with ray-tracing results, refer to[7].

1. As it can be seen from Fig. 1, an off-axis ray impinges the mirror shell at two different angles, $\alpha_1$ and $\alpha_2$, which in general depend on the impact position. In double cone approximation, indeed, they depend solely on the polar angles $\varphi_1$, $\varphi_2$, of impact on the two segments, measured from the off-axis plane. If $\theta$ is a small angle, $\varphi_1 \approx \varphi_2$ and we denote with $\varphi$ their common value. In these conditions, the incidence angles have the expressions

$$\alpha_1(\varphi) = \alpha_0 + \delta - \vartheta \cos\varphi, \quad (1)$$

$$\alpha_2(\varphi) = \alpha_0 - \delta + \vartheta \cos\varphi. \quad (2)$$

2. Denoting with $r_\lambda(\alpha)$ the reflectivity of the mirror coating at the incidence angle $\alpha$ and the X-ray wavelength $\lambda$, the mirror effective area at $\lambda$ in double cone approximation is

$$A_D(\lambda,\theta) = 2R_0 \int_0^\pi (L\alpha)_{min} r_\lambda(\alpha_1) r_\lambda(\alpha_2)\, d\varphi, \quad (3)$$

where $(L\alpha)_{min}$= min($L_1\alpha_1$, $L_2\alpha_2$) if positive, otherwise $(L\alpha)_{min}$= 0.

3. In the frequent case $L_1 = L_2 = L$, Eq. 3 reduces to

$$A_D(\lambda,\theta) = 2R_0 L \int_0^{\pi} \alpha_{\min} r_\lambda(\alpha_1) r_\lambda(\alpha_2)\, d\varphi, \quad (4)$$

where $\alpha_{\min}$ is the smallest between $\alpha_1$ and $\alpha_2$, and has a compact expression:

$$\alpha_{\min}(\varphi) = \max(0, \alpha_0 - |\delta - \vartheta\cos\varphi|), \quad (5)$$

4. For a source at infinity and $L_1 = L_2$, Eq. 4 reduces simply to

$$A_\infty(\lambda,\theta) = 4R_0 L \int_0^{\pi/2} \alpha_1 r_\lambda(\alpha_1) r_\lambda(\alpha_2)\, d\varphi, \quad (6)$$

where $\alpha_1 = \alpha_0 - \theta\cos\varphi$ if positive, and zero otherwise, is $\alpha_{\min}$ in the first quadrant, and $\alpha_2 = \alpha_0 + \theta\cos\varphi$. On axis, $\theta = 0$ and Eq. 6 simply becomes

$$A_\infty(\lambda,0) = 2\pi R_0 L \alpha_0 r_\lambda^2(\alpha_0). \quad (7)$$

5. In the ideal case of an ideal reflectivity $r_\lambda(\alpha) = 1$, Eq. 6 can be solved explicitly to obtain the geometric area, $A_\infty(\theta)$, of the mirror with $L_1 = L_2$ for a source at infinity, for off-axis angles $\theta < \alpha_0$:

$$A_\infty(\theta) = A_\infty(0)\left(1 - \frac{2\theta}{\pi\alpha_0}\right). \quad (8)$$

Eq. 8 is exactly the result found empirically by Van Speybroeck and Chase in 1972[8] for the off-axis vignetting of a Wolter-I mirror, excepting the approximation $\pi \approx 3$.

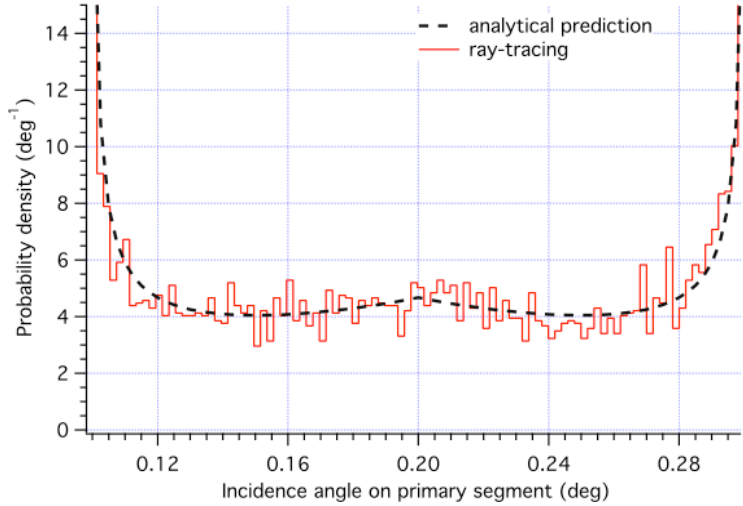

Fig. 2: the incidence angle distribution on the primary segment of a Wolter-I mirror shell with $\alpha_0 = 0.2$ deg and off-axis by $\theta = 0.1$ deg, as resulting from a ray-tracing (histogram), and the same distribution obtained analytically (line) by plotting the normalized distribution reported in Eq. 4, for $\alpha_1 < \alpha_0$.

The applicability of the mentioned double-cone approximation to Wolter-I mirrors increases with the focal length: more exactly, the error we commit in the effective area computation is of the order of $L/f$ or better, therefore it is of the order of a few percent in real cases. For a source at infinity and if $L_1 = L_2$, if $\theta \neq 0$, we obtain an alternative form of the Eq. 6, along with a change in the integration variable from $\varphi$ to $\alpha_1$,

$$A_\infty(\lambda,\theta) = 4R_0 L \int_{\alpha_m}^{\alpha_0} \frac{\alpha_1 r_\lambda(\alpha_1) r_\lambda(\alpha_2)}{\sqrt{\theta^2 - (\alpha_1 - \alpha_0)^2}}\, d\alpha_1, \quad (9)$$

where $\alpha_m$ = max(0, $\alpha_0 - \theta$). We note that such a change of variable would not be possible if $\theta = 0$, because the function $\alpha_1 = \alpha_0 - \theta \cos\varphi$ would collapse to a constant. The effective area expression in Eq. 9, when put in this form, has the advantage that the integration is directly performed over the incidence angle range in which the reflectivity is calculated. For this reason, this form is suitable to compute the effective area as a function of $\theta$, at a single X-ray wavelength $\lambda$. In contrast, Eq. 6 is better suited for computing $A_D(\lambda, \theta)$ at fixed $\theta$.

It is worth noting that Madsen et al.[9] independently found an expression similar to Eq. 9, but the weight function, $W_{inc}$, that appeared in there was not written explicitly because it was derived from a ray tracing. Comparing that equation with Eq 9, we can now write the explicit expression of $W_{inc}$, adapting the notation to the one in use in this work,

$$W_{inc}(\alpha_0, \alpha_1) = \frac{2\alpha_1}{\pi \alpha_0 \sqrt{\theta^2 - (\alpha_1 - \alpha_0)^2}}, \tag{10}$$

provided that the integration is performed only in the region where $\alpha_1 \leq \alpha_0$. Where $\alpha_1 > \alpha_0$, the same distribution is valid, after substituting $\alpha_1 \to \alpha_2$: the resulting curve is thereby symmetric with respect to $\alpha_0$. The $W_{inc}$ function, which should be interpreted as an effective area density per incidence angle unit, is plotted in Fig. 2, after normalization. The area distribution obtained from a ray tracing is also shown, finding an excellent agreement. The angle distribution of Eq. 10 obtained is extensively used in another paper of this volume[10].

The off-axis effective area of the optical module is simply obtained by summing the off-axis effective area of all mirrors, on condition that the mutual obstruction of mirrors is negligible. If the mirrors are densely nested, the total effective area is smaller than the sum of the effective area of individual mirrors. However, in the following we suppose that the obstruction off-axis is negligible.

The expression of the effective area as a function of the off-axis angle also allows to measure the field of view of the optical module, $F(\lambda)$, usually defined as the angular diameter at which the effective area is halved with respect to the on-axis value. $F(\lambda)$ is also related to another very important quantity: the *grasp*, $G(\lambda)$, defined as the product of the effective area on-axis and the solid angle delimited by the field of view,

$$G(\lambda) \approx \frac{\pi}{4} F^2(\lambda) A_\infty(\lambda, 0). \tag{11}$$

Such a definition is not very rigorous, because it supposes that the effective area changes slowly for small off-axis angles $\theta$ within $F(\lambda)$ with respect to $A_\infty(\lambda, 0)$, then that it suddenly drops outside the field of view. A more general definition can be

$$G_\Theta(\lambda) \approx \int A_\infty(\lambda, \theta) \, d\Omega, \tag{12}$$

where $d\Omega = 2\pi \sin\theta \, d\theta$ is the infinitesimal solid angle centered on the optical axis, and the integration is extended up to a maximum off-axis angle $\Theta$, where the effective area can be considered negligible. The dependence on $\lambda$, the X-ray wavelength, entirely comes from the $\lambda$ dependence of the coating reflectivity. We note that the grasp is additive, therefore Eq. 12 applies to single mirrors as well as non-obstructed mirror assemblies. Moreover we note that, according this definition, the on-axis area does not contribute to the grasp, because for $\theta = 0$ the integrand vanishes.

When applied to a single mirror with primary segment slope $\alpha_0$, the integral in Eq. (12) is not of immediate computation because it requires a double integration: the first one in $\varphi$ and the second one in $\theta$. However, we can reduce the computation to a single integration. In fact, it can be shown that the grasp of a single mirror can be computed directly from the mirror reflectivity of the coating:

$$G_\Theta(\lambda) \approx 8\pi R_0 L \int_0^{\alpha_0} \alpha_1 \sqrt{\Theta^2 - (\alpha_1 - \alpha_0)^2} \, r_\lambda(\alpha_1) r_\lambda(\alpha_2) \, d\alpha_1 \tag{13}$$

The proof of the Eq. 13 is reported in Appendix. We note that the grasp explicitly depends on the maximum off-axis angle $\Theta$, even if it cannot be much larger than $\alpha_0$ in order to avoid the obstruction from neighboring shells.

## 3. THE OFF-AXIS EFFECTIVE AREA OF THE NHXM MODULES

In this section we present the application of the formulae reported in Sect. 2 to the computation of the effective area of the NHXM hard X-ray telescope[2]. The NHXM optical system, as resulting from a detail optimization[11], comprises 4 identical modules with 70 mirror shells of 391 to 155 mm diameter and a 10 m focal length. The $\alpha_0$ angle takes on values in the interval 0.28 – 0.11 deg, and $L_1 = L_2 = 300$ mm for all shells; therefore the effects of the deviation from the double cone profile on the effective area can be computed to be < $L/f$ = 3%, and the formulae for the effective area reported in Sect. 2 can be applied to a good approximation. The multilayer coatings are Pt/C of the graded type, in order to effectively reflect X-rays in a wide energy band, and in particular the shells with the shallowest angles have to reflect up to 80 keV. To obtain such a reflectance with the minimum number of bilayers, the shells are collected into 9 groups, and for each of them a specific multilayer recipe[11] has been optimized. In this work we assume that configuration as a baseline for the NHXM optical modules.

However, before adding up the effective area of mirrors to obtain the effective area of the modules, another point has to be checked: the mutual obstruction of shells has to be negligible over the requested telescope's field of view, 12 arcmin diameter. To this end, a 6 arcmin angular aperture – as seen from the intersection plane – is left between each couple of adjacent shells. Considering that the thickness of the mirror walls varies from 0.34 to 0.13 mm, going inwards, this implies that a gap of 1.97 to 1.1 mm, always going inwards, has to be left between shells at the intersection plane. Actually, this is certainly sufficient to avoid the obstructions up for $\theta$ < 5 arcmin, whilst for larger off-axis angles there is some effective area loss due to the impact of some rays on the outer surface of the secondary segments after the second reflection. Nevertheless, it is possible to compute that such effective area loss at $\theta$ = 6 arcmin is limited to a few percent. Therefore, we hereafter assume the equations of Sect. 2 as valid within a few percents accuracy.

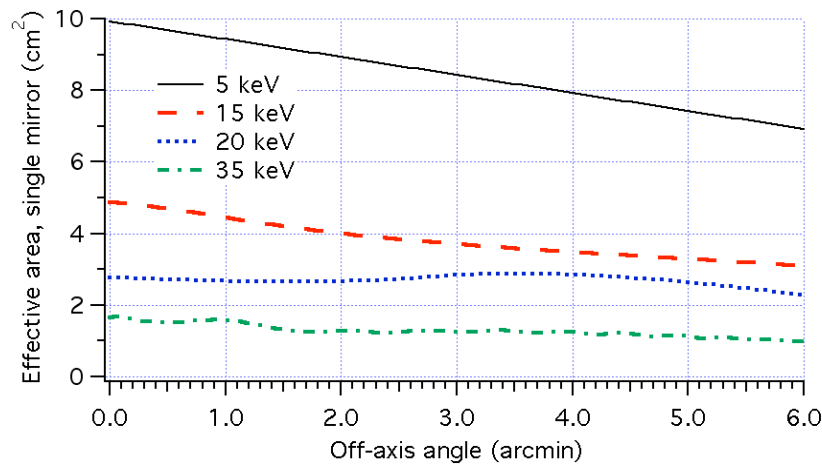

Fig. 3: the effective area of a Wolter-I mirror with $2R_0$ = 297 mm, $L$ = 300 mm, $f$ = 10 m, adopting as a multilayer recipe[11] a 169 couples of Pt/C layers, minimum d-spacing 24.7 Å, maximum d-spacing 146.7 Å, a power-law exponent 0.223, and a thickness ratio $\Gamma$ = 0.42 as a function of the off-axis angle, for different X-ray energies. A 4 Å roughness rms is assumed.

As a first example of computation, we show the effective area of a single mirror, as a function of the off-axis angle, for selected X-ray energies (Fig. 3): the mirror and the multilayer coating parameters are also reported therein. The effective area values are computed using Eq. 9. We note that at 5 keV the effective area variation essentially follows a linear decrease, in accord with Eq. 8, because the coating reflectivity at that energy is nearly constant with the incidence angle; therefore the effective area trend is almost indistinguishable from the decrease of the sole geometric area. At higher energies, the effective area also decreases with the off-axis angle, but with oscillating trends depending on the energy chosen, because of the oscillatory behavior of the multilayer reflectivity.

The second computation example regards the total effective area of a module of NHXM, which can be calculated as a function of the X-ray energies for some values of $\theta$. The result is displayed in Fig. 4, on-axis, at 2, 4, and 6 arcmin off-axis. The effective area values up to 80 keV fulfills the requirement of a > 1000 cm$^2$ at 1 keV (and even the 1500 cm$^2$ goal), 350 cm$^2$ at 30 keV, and 100 cm$^2$ at 70 keV on-axis, with the three imaging modules.

Moreover, from the decrease of the effective area with the off-axis angle we can evaluate the Field of View of the module: > 12 arcmin up to 40 keV, and decreasing as the X-ray energy is increased: ~9 arcmin at 50 keV, ~4 arcmin at 70 keV. However, the exact value cannot be computed accurately because a much larger sampling of the off-axis angle would be needed.

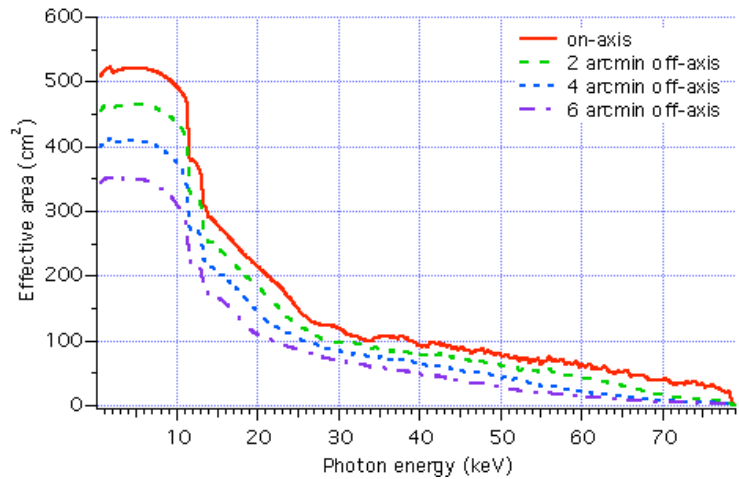

Fig. 4: the effective area as a function of the X-ray energy, computed using Eq. 6, for a single module of NHXM, with optimized Pt/C multilayer recipes, assuming a 4 Å roughness rms and a 10% vignetting due to the spider obstruction. The different curves are traced for various off-axis angles.

As a last example, we show in Fig. 5 the variation of the *total* effective area of a single module of the NHXM telescope, as a function of the off-axis angle, for some selected X-ray energies. The computation has been performed in a fast way by applying Eq. 9 to each mirror in the module and summing the curves obtained for all mirrors. The result is similar to Fig. 3, with a linear decrease at 5 keV essentially determines by the sole geometric vignetting, and with different slopes at higher energies determined by the chosen multilayer recipes. In this case, however, the oscillatory trends seen in the case of a single mirror are not present because they are averaged out when the contribution of the individual mirrors are summed up. The field of view, in this case, can be accurately measured for all the energies listed in Fig. 5. At this regard, we note that all effective areas are not halved yet at 6 arcmin, the radius of the NHXM field of view, and that the effective area curves do not exhibit gaps, peaks or dips. This is a confirmation of the good performance of the adopted set of multilayer recipes. The effects of the multilayer optimization on the field of view are treated with more detail in another conference proceeding of this volume[10].

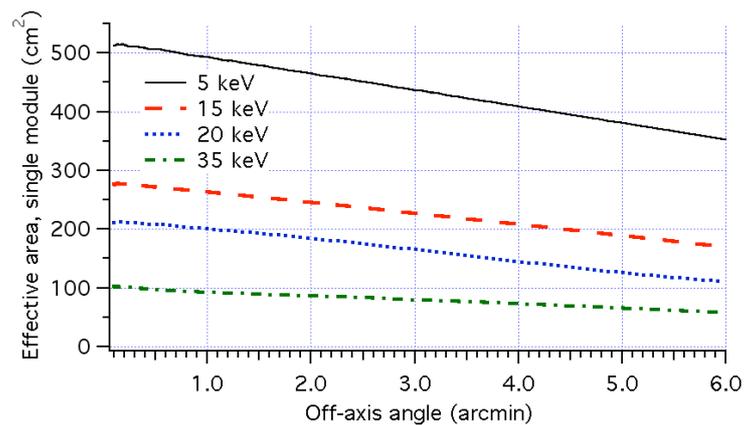

Fig. 5: the effective area as a function of the off-axis angle, computed using Eq. 9, for a single module of NHXM, with optimized Pt/C multilayer recipes, at selected X-ray energies. Note that at all considered energies the effective area at 6 arcmin off-axis is more than half the on-axis value, therefore the optics' field of view is wider than 12 arcmin diameter.

## 4. CONCLUSIONS

In this work we have presented the application to the New Hard X-ray mission of a set of novel, analytical formulae to compute the off-axis effective area of a Wolter-I X-ray mirror in double cone approximation, and more generally to compute the off-axis effective area of an X-ray optical module. We have also obtained some derived formulae to compute quickly the field of view and the grasp of a grazing incidence X-ray mirror. The method can be applied to any reflective coating; therefore it will be of considerable usefulness in future optimizations of the optical design of grazing incidence X-ray optical modules.

## APPENDIX: INTEGRAL FORMULA FOR THE GRASP OF AN X-RAY MIRROR

In this appendix we reduce the grasp formula for a single mirror (Eq. 12) to a single integration over the incidence angle on the primary segment of the mirror, $\alpha_1$. Using the definition of $d\Omega = 2\pi \sin\theta \, d\theta$, we write Eq. 12 as

$$G_\Theta(\lambda) \approx 2\pi \int_0^\Theta A_\infty(\lambda,\theta) \sin\theta \, d\theta \tag{A1}$$

where $\Theta$ is the maximum off-axis angle to which we extend the computation. Because $0 < \theta < \Theta$ is shallow, we can approximate $\sin\theta \sim \theta$. Using Eq. 9, we can rewrite the expression for the grasp as

$$G_\Theta(\lambda) \approx 8\pi R_0 L \int_0^\Theta \theta \, d\theta \int_{\alpha_m}^{\alpha_0} \frac{\alpha_1 r_\lambda(\alpha_1) r_\lambda(\alpha_2)}{\sqrt{\theta^2 - (\alpha_1 - \alpha_0)^2}} \, d\alpha_1. \tag{A2}$$

where $\alpha_m = \max(0, \alpha_0 - \theta)$. For a fixed $\alpha_1$, the $\theta$ variable takes on values in the interval $\alpha_0 - \alpha_1 < \theta < \Theta$ (see Fig. 6). We thereby derive, by exchanging the integration order,

$$G_\Theta(\lambda) \approx 8\pi R_0 L \int_0^{\alpha_0} \alpha_1 r_\lambda(\alpha_1) r_\lambda(\alpha_2) \, d\alpha_1 \int_{\alpha_0 - \alpha_1}^\Theta \frac{\theta \, d\theta}{\sqrt{\theta^2 - (\alpha_1 - \alpha_0)^2}}. \tag{A3}$$

The second integral can be immediately solved, and we derive Eq. 13:

$$G_\Theta(\lambda) \approx 8\pi R_0 L \int_0^{\alpha_0} \alpha_1 \sqrt{\Theta^2 - (\alpha_1 - \alpha_0)^2} \, r_\lambda(\alpha_1) r_\lambda(\alpha_2) \, d\alpha_1. \tag{A4}$$

We note that, unlike in Eq. 9, in Eq. A4 the integrand is finite everywhere.

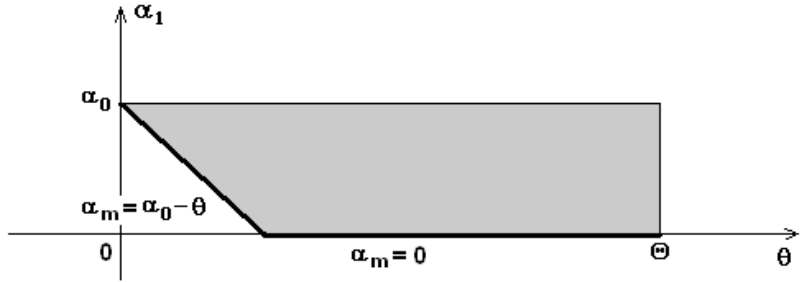

Fig. 6: the integration set (the grayed area) in the $(\theta, \alpha_1)$ space.

For the ideal case of $r_\lambda(\alpha) = 1$ for any $\alpha$ and $\lambda$, and if $\Theta = \alpha_0$, Eq. A4 becomes

$$G_{\alpha_0} \approx 8\pi R_0 L \int_0^{\alpha_0} \alpha_1 \sqrt{\alpha_1 \alpha_2} \, d\alpha_1, \tag{A5}$$

that, after some algebra, becomes

$$G_{\alpha_0} \approx 2\pi\left(\pi - \frac{4}{3}\right)R_0 L \alpha_0^3. \tag{A6}$$

The same exact result can be obtained by direct substitution of the geometric area expression (Eq. 8) into the definition of grasp (Eq. 12). Assuming also Eq. 11 as valid, we also obtain the geometric field of view,

$$F \approx 2\alpha_0\sqrt{1 - \frac{4}{3\pi}} \approx 1.517\alpha_0, \tag{A7}$$

while by setting Eq, 8 to half the area on-axis we obtain $F = \pi\alpha_0/2 \approx 1.57\alpha_0$, in quite good agreement with Eq. A7.

## ACKNOWLEDGMENTS

This work has been financed by the Italian Space Agency (contract I/069/09/0).

## REFERENCES

[1]. J. E. Koglin, H. An, K. L. Blaedel, N. F. Breinholt, F. E. Christensen, W. W. Craig, T. A. Decker, C. J. Halley, L.C. Hale, F. A. Harrison, C. P. Jensen, K. K. Madsen, K. Mori, M. J. Pivovaroff, G. Tajri, W. W. Zhang, "NuSTAR hard X-ray optics design and performance". *SPIE Proc.* **7437**, 74370C (2009)
[2]. S. Basso, G. Pareschi, O. Citterio, D. Spiga, G. Tagliaferri, M. Civitani, L. Raimondi, G. Sironi, V. Cotroneo, B. Negri, G. Parodi, F. Martelli, G. Borghi, A. Orlandi, D. Vernani, G. Valsecchi, R. Binda, S. Romaine, P. Gorenstein, P. Attina', "The optics system of the New Hard X-ray Mission: design and development", *SPIE Proc.* **7732**, this conference (2010)
[3]. A. Furuzawa, Y. Ogasaka, H. Kunieda, T. Miyazawa, M. Sakai, Y. Kinoshita, M. Makinae, S. Sasaya, Y. Kanou, D. Niki, T. Ohgi, N. Oishi, K. Yamane, N. Yamane, N. Yishida, Y. Haba, Y. Tawara, K. Yamashita, M. Yishida, Y. Maeda, H. Mori, K. Tamura, H. Awaki, and T. Okajima, "The current status of the ASTRO-H/HXT development facility", *SPIE Proc.* **7437**, 743709 (2009)
[4]. IXO web page, http://ixo.gsfc.nasa.gov/
[5]. V. Cotroneo, P. Conconi, G. Cusumano, G. Pareschi, D. Spiga and G. Tagliaferri, "Effects of small oscillations on the Effective Area", *AIP conference proceedings*, CP1126, 88-90 (2009)
[6]. G. Pareschi, P. Attina', S. Basso, G. Borghi, W. Burkert, R. Buzzi, O. Citterio, M. Civitani, P. Conconi, V. Cotroneo, G. Cusumano, E. Dell'Orto, M. Freyberg, G. Hartner, P. Gorenstein, E. Mattaini, F. Mazzoleni, G. Parodi, S. Romaine, D. Spiga, G. Tagliaferri, R. Valtolina, G. Valsecchi, D. Vernani, "Design and development of the SIMBOL-X hard X-ray optics", *SPIE Proc.* **7011**, 70110N (2008)
[7]. D. Spiga, V. Cotroneo, S. Basso, P. Conconi, "Analytical computation of the off-axis effective area of grazing-incidence X-ray mirrors", *Astronomy and Astrophysics* 505, 373-384 (2009)
[8]. L. P. Van Speybroeck, R. C. Chase, "Design Parameters of Paraboloid - Hyperboloid Telescopes for X-ray astronomy", Appl. Opt. 11(2), 440-445 (1972)
[9]. K. K. Madsen, F. A. Harrison, P. H. Mao, F. E. Christensen, C. P. Jensen, N. Brejnholt, J. Koglin, M. J. Pivovaroff, "Optimization of Pt/SiC and W/Si multilayers for the Nuclear Spectroscopic Telescope Array", *SPIE Proc.* **7437**, 743716 (2009)
[10]. V. Cotroneo, G. Pareschi, D. Spiga, G. Tagliaferri, "Effects of the coating optimization on the field of view of a Wolter X-ray telescope", *SPIE Proc.* **7732**, this conference (2010)
[11]. V. Cotroneo, G. Pareschi, D. Spiga, G. Tagliaferri, "Optimization of the reflecting coatings for the New Hard X-ray Mission", *SPIE Proc.* **7437**, 743717 (2009)